\documentclass[sigconf]{acmart}

\usepackage{amsfonts}
\usepackage{dsfont}

\usepackage[subrefformat=parens]{subcaption}
\usepackage{wrapfig}
\usepackage{enumitem}
\usepackage{lipsum}
\usepackage{mathtools}

\copyrightyear{2021}
\acmYear{2021}
\setcopyright{acmcopyright}
\acmConference[KDD '21] {Proceedings of the 27th ACM SIGKDD Conference on Knowledge Discovery and Data Mining}{August 14--18, 2021}{Virtual Event, Singapore.}
\acmBooktitle{Proceedings of the 27th ACM SIGKDD Conference on Knowledge Discovery and Data Mining (KDD '21), August 14--18, 2021, Virtual Event, Singapore}
\acmPrice{15.00}
\acmISBN{978-1-4503-8332-5/21/08}
\acmDOI{10.1145/3447548.3467248}

\setlist[itemize]{leftmargin=*}
\definecolor{blue(pigment)}{rgb}{0.2, 0.2, 0.6}

\makeatletter
\newcommand\footnoteref[1]{\protected@xdef\@thefnmark{\ref{#1}}\@footnotemark}
\makeatother

\begin{document}
\fancyhead{}

\title{Dynamic Hawkes Processes for Discovering Time-evolving Communities' States behind Diffusion Processes}


\author{Maya Okawa$^{1, 3}$, Tomoharu Iwata$^2$, Yusuke Tanaka$^2$, Hiroyuki Toda$^1$ \and Takeshi Kurashima$^1$, Hisashi Kashima$^3$}
\affiliation{
  \institution{$^1$NTT Service Evolution Labs., NTT Corporation}
  \city{}
  \country{}
}
\email{{maya.ookawa.af,hiroyuki.toda.xb,takeshi.kurashima.uf}@hco.ntt.co.jp}
\affiliation{
  \institution{$^2$NTT Communication Science Labs., NTT Corporation}
  \city{}
  \country{}
}
\email{{tomoharu.iwata.gy,yusuke.tanaka.rh}@hco.ntt.co.jp}
\affiliation{
  \institution{$^3$Department of Intelligence Science and Technology, Kyoto University}
  \city{}
  \country{}
}
\email{kashima@i.kyoto-u.ac.jp}

\begin{abstract}
Sequences of events including infectious disease outbreaks, social network activities, and crimes are ubiquitous and the data on such events carry essential information about the underlying diffusion processes between communities (e.g., regions, online user groups). Modeling diffusion processes and predicting future events are crucial in many applications including epidemic control, viral marketing, and predictive policing. Hawkes processes offer a central tool for modeling the diffusion processes, in which the influence from the past events is described by the triggering kernel. However, the triggering kernel parameters, which govern how each community is influenced by the past events, are assumed to be static over time. In the real world, the diffusion processes depend not only on the influences from the past, but also the current (time-evolving) states of the communities, e.g., people’s awareness of the disease and people’s current interests. In this paper, we propose a novel Hawkes process model that is able to capture the underlying dynamics of community states behind the diffusion processes and predict the occurrences of events based on the dynamics. Specifically, we model the latent dynamic function that encodes these hidden dynamics by a mixture of neural networks. Then we design the triggering kernel using the latent dynamic function and its integral. The proposed method, termed DHP (Dynamic Hawkes Processes), offers a flexible way to learn complex representations of the time-evolving communities' states, while at the same time it allows to computing the exact likelihood, which makes parameter learning tractable. Extensive experiments on four real-world event datasets show that DHP outperforms five widely adopted methods for event prediction. 
\end{abstract}



\maketitle

\section{Introduction}
Various social phenomena can be described by diffusion processes among multiple communities. For example, infectious diseases like COVID-19 are transmitted from one county to another, leading to a worldwide pandemic \cite{tatem2006global}. 
Information such as opinions, news, and articles are shared and disseminated among online communities,  
e.g., user groups in social networks, news websites, and blogs. 
Such diffusion phenomena are recorded as multiple sequences of events, which indicate when and in which community the event occurred.  Understanding the diffusion mechanism and predicting future events are crucial for many practical applications across domains.
For example, policymakers would be able to design prompt and appropriate interventions to curb the spread of disease  
given a better understanding of the mechanisms behind the transmission and more reliable predictions.  

Temporal point processes provide an elegant mathematical framework for modeling event sequences.  
In these methods, the probability of event occurrences is determined by the {\sl intensity} function. 
Hawkes process is an important class of point processes for modeling diffusion processes. 
These models use the {\sl triggering kernel} to characterize diffusion processes and estimate its parameters via maximum likelihood.  
The triggering kernel encodes the magnitude and speed of influence from the past events, namely, 
how likely and quickly the past events in one community (i.e., ``source'' community) will affect the occurrence of a particular event in another community (i.e., ``target'' community). 
Hawkes process and its variants have been applied in diverse areas,  
from epidemic modeling \cite{kim2019modeling} to social network analysis \cite{zhao2015seismic,rizoiu2017hawkes,farajtabar2017coevolve}. 
However, they have focused on learning the static influence of the past events on the current event, 
thereby largely overlooking the factor of time-evolution. 
In reality, the diffusion processes depend not only on the influences from the past but also on the current state of the target communities. 
For example, the outbreaks of infectious diseases in one community (e.g., country) can also be driven by people's awareness of the disease in each community (country) and their preventive behaviors  
which can constantly change over time, on top of the past record of the disease occurrence.
As another example, information diffusion heavily depends on ongoing peoples' interests in the target community (e.g., online user group).  
In particular, the spread of information to one target community (user group) is strengthened when an topic deemed to be important by the target community emerges in the online space; 
while it is weakened in accordance with a gradual loss in peoples' interest in the topic.

A few studies have considered the underlying dynamics of such ``states'' in communities. For instance, the SIR-Hawkes model \cite{rizoiu2018sir} redesigned the triggering kernel of the Hawkes process by incorporating the recovered (immune) population dynamics 
over the course of the pandemic. 
Kobayashi {\sl et al.} \cite{kobayashi2016tideh} proposed a time-dependent triggering kernel that varies periodically in time 
for modeling daily cycles of human activity. 
However, these approaches rely on hand-crafted functions for describing the latent dynamics of states and so demand expert domain knowledge. Moreover, they may not be flexible enough to accommodate the complexity and heterogeneity of the real world.
In fact, in many practical applications  
the complete set of factors is largely unknown and thus difficult to model through restricted parametric forms. 
Taking information diffusion as an example, the time-evolution of peoples' interest on a given topic is generally unknown and not directly observable. 

A potential solution is to directly model the triggering kernel parameters using a flexible function of time (e.g., neural network). 
Alas, naively employing this approach makes parameter learning intractable since the log-likelihood of Hawkes processes involves the integral of the triggering kernel. 
Computing the integral of the triggering kernel in combination with the neural network is generally infeasible. 

In this paper, we propose a novel Hawkes process model referred to as \textsf{DHP} (Dynamic Hawkes Process) 
which automatically learns the underlying dynamics of the communities' states behind the diffusion processes in a manner that allows tractable learning. 
We introduce the {\sl latent dynamics function} for each community that represents its hidden dynamic states. 
Our core idea is to extend the triggering kernel by combining it with the latent dynamics function and its integral.  
Specifically, we model the magnitude of diffusion by the latent dynamics function and the speed of diffusion by the integral of the latent dynamics function.  
This design choice offers two benefits.  
First, the resulting triggering kernel can be expressed as a product of two components: 
composite function with the ``inner'' function being the integral of the latent dynamics function and the ``outer'' function being the basic triggering kernel;  
and the derivative of the inner function of that composite function (i.e., the latent dynamics function). 
Hence, by applying the substitution rule for definite integrals (i.e., the chain rule in reverse),  
we can obtain a closed-form solution for the integral of the triggering kernel involved in the log-likelihood.  
Second, it allows capture of the simultaneous changes of magnitude and speed of the diffusion 
as they are related through the latent dynamics function and its integral,  
which is desirable for many applications. 
For example, in the context of disease spread, active preventitive measures can reduce both the magnitude and the speed of the infection.
To model the integral of the latent dynamics function,  
we utilize and extend a monotonic neural network \cite{sill1998monotonic,chilinski2020neural}.  
This formulation enables \textsf{DHP} to learn flexible representations of the community state dynamics that underlie the diffusion processes. 
It should be noted that \textsf{DHP} can be easily extended to capture the time-evolving relationships between communities, by introducing the latent dynamics function 
for pairs of communities.  
In this work, we adopt \textsf{DHP} to demonstrate the hidden state dynamics of individual communities.

\begin{figure}[t]
    \hspace{-5mm}\centering\includegraphics[width=0.83\linewidth]{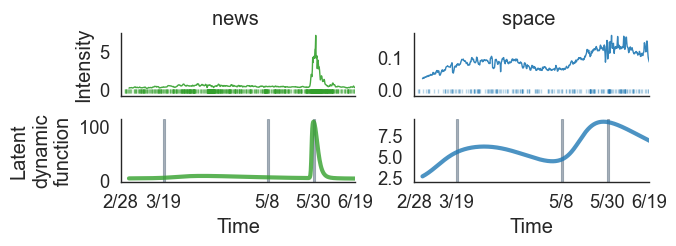}
    \centering\includegraphics[width=0.87\linewidth]{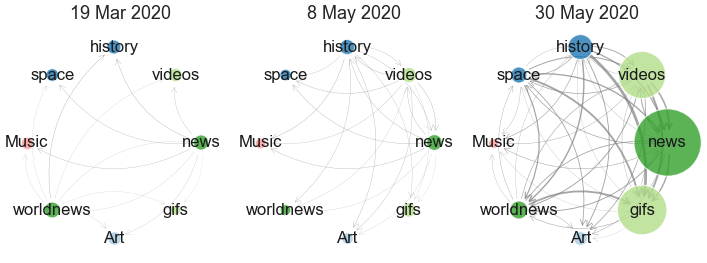}
\vspace{-2mm}
\caption{
Dynamics of community states learned by our \textsf{DHP} for Reddit dataset. 
Top: Intensity with observed event sequences and latent dynamics function for two Reddit communities (i.e., subreddits): \texttt{news} and \texttt{space}. 
Bottom: Learned triggering kernel between 8 selected subreddits at 3 different time points. 
Nodes denote subreddits, color indicates category.  
Node size is proportional to latent dynamics function for each subreddit. 
Edge width is proportional to triggering kernel, which indicates strength of diffusion between pairs of subreddits. 
We can see the latent dynamics function increases over time for most subreddits from March to May, 2020. 
It increases rapidly for \texttt{news} and slowly for \texttt{space} following the onset of the COVID-19 lockdown.  
Our \textsf{DHP} automatically learns how activities of each community evolve over the course of the pandemic. 
}\label{fig:reddit_five_sample}
\vspace{-9mm}
\end{figure}

The main contributions of this paper are as follows:
\begin{itemize}
\item We propose a novel Hawkes process framework, \textsf{DHP} (Dynamic Hawkes Process) for modeling diffusion processes and predicting future events. 
The proposal, \textsf{DHP}, is able to learn the time-evolving dynamics of community states behind the diffusion processes.
\item We introduce {\sl latent dynamics function}; it reflects the hidden community dynamics and design the triggering kernel of the Hawkes process intensity using the latent dynamics function and its integral.  
The resulting model is computationally tractable and flexible enough to approximate the true evolution of the community states underlying the diffusion processes.  
\item We carry out extensive experiments using four real-world event datasets: Reddit, News, Protest, and Crime. 
The results show that \textsf{DHP} outperforms the existing works. 
Case studies demonstrate that \textsf{DHP} uncovers the hidden state dynamics of communities which underlie the diffusion processes 
by the latent dynamic function (See Figure \ref{fig:reddit_five_sample}).  
\end{itemize}

\section{Related Work}
With the evolution of data collection technology, 
extensive event sequences with precise timestamps are becoming available in an array of fields  
such as public health and safety \cite{mohler2011self,weiss2013forest,lasko2014efficient}, 
economics and finance \cite{chavez2005estimating,bacry2015hawkes,hawkes2018hawkes}, 
communications \cite{farajtabar2017coevolve}, 
reliability \cite{baxter1996point,veber2008generalized,tanwar2014imperfect}, and  
seismology \cite{ogata1988statistical,ogata1984inference,ogata1999seismicity}. Temporal point processes provide a principled theoretical framework for modeling such event sequences,  
in which the occurrences of events are determined by the intensity function.  

Classical examples of temporal point processes include 
reinforced Poisson process \cite{pemantle2007survey}, self-correcting point process \cite{isham1979self}, 
and Hawkes process \cite{hawkes1971spectra}. 
The Reinforced Poisson process \cite{pemantle2007survey} considers the cumulative count of past events  
and a time-decreasing trend, and has been recently applied for predicting online popularity \cite{shen2014modeling}.  
The intensity of the self-correcting point process \cite{isham1979self} increases steadily and this trend is corrected by past observed events. 
Although these models have been widely used, they are not suitable for modeling diffusion processes between communities  
as they cannot explicitly model the influence of the past events underlying diffusion processes. 
Hawkes process \cite{hawkes1971spectra} explicitly models the influence of the past events and captures triggering patterns between events (i.e., diffusion processes).  
Hawkes processes have been proven effective for modeling diffusion processes, including 
earthquakes and aftershocks \cite{mohler2011self}, near-repeat patterns of crimes \cite{mohler2011self}, 
financial transactions \cite{embrechts2011multivariate,bacry2015hawkes,hawkes2018hawkes}, 
online purchases \cite{xu2014path,du2015time,dai2016deep,wang2016coevolutionary}, 
and information cascades \cite{zhao2015seismic,rizoiu2017hawkes,farajtabar2017coevolve}.  

Recent studies employ neural network architectures to model point process intensity.
In \cite{du2016recurrent}, the authors design the intensity using RNN. 
Omi {\sl et al.} \cite{omi2019fully} extended our work by combining it with a monotonic neural network. 
Compared to classical point process methods, RNN-based models provide a more flexible way to handle the complex dependencies between events. 
However, the above methods focus on learning the triggering patterns of diffusion processes, i.e., influences from past events, and disregard the current (time-evolving) states of the communities.  
In the real world, event occurrences largely depend on the current community states (e.g., people’s awareness of the disease, ongoing people’s interest), 
which can evolve over time, as well as the past.

Several studies incorporate the time-variant dynamics of the community states behind diffusion processes into the Hawkes process formulation.  
For instance, the SIR-Hawkes model \cite{rizoiu2018sir} considers recovered (immune) population dynamics 
to enhance the prediction of infectious disease events over the course of pandemic.  
Kobayashi {\sl et al.} \cite{kobayashi2016tideh} proposed a time-dependent Hawkes process that accounts for the circadian and weekly cycles of human activity. 
Navaroli {\sl et al.} \cite{navaroli2015modeling} used nonparametric estimation to learn cyclic human activities underlying digital communications. 
All of the above methods, unfortunately, rely on a domain expert's knowledge to elucidate the 
dynamics of the communities states behind diffusion processes. 
Such dynamics are often quite complex and remain unexplored in many practical applications.

Different from the existing methods, our proposed method 
both incorporates the temporal dynamics of communities’ states and the past influences.

\section{Preliminaries}
This section provides the general framework of point processes on which our work is built,  
and the formal definition of the event prediction problem studied in this paper.  

\subsection{Hawkes Processes}
Point process is a random sequence of events occurring in continuous time $\{t_1, t_2, \cdots t_I\}$, with $t_i\in[0,T)$.  
Point processes are fully determined by ``intensity'' function $\lambda(t)$.
Given the history of events $\mathcal{H}(t)$ up to time $t$, the intensity is defined as  
\begin{align}
\lambda(t) \equiv \lim_{\Delta t\rightarrow 0} \frac{\mathbb{E}[N(t+\Delta t)-N(t)|\mathcal{H}(t)]}{\Delta t},
\end{align}
where $N(t)$ is a number of events falling in $[0,t)$, $\Delta t$ is a small time interval, and $\mathbb{E}$ is an expectation.   
The intensity value $\lambda(t)$ at time $t$ measures the probability that an event occurs in the infinitesimal time interval $[t, t+\Delta t)$
given past events $\mathcal{H}(t)$.  

Hawkes process \cite{hawkes1971spectra} is an important class of point processes, and can describe self-exciting phenomena. 
The intensity of Hawkes process is defined as  
\begin{align}
\lambda(t) = \mu + \sum_{j: t_j<t} g(t-t_j),  
\end{align}
where $\mu\geq 0$ is a background rate and $g(\cdot)\geq 0$ is a {\sl triggering kernel}  encoding the augmenting or attenuating effect of past events on current events.  
Intuitively, each event at time $t_j$ elevates the occurrence rate of events at time $t$ by the amount $g(t-t_j)$ for $t>t_j$. 

The univariate Hawkes process can be extended to multivariate Hawkes Process (MHP) to handle the mutual excitation of events (i.e., diffusion) among different communities (denoted by dimensions). Suppose we have $I$ historical observations $\mathcal{D}=\{(t_i,m_i)\}_{i=1}^I$  
with time $t_i\in [0,T)$ and community $m_i\in \{1,...,M\}$. 
In our setting, the communities indicate countries, city districts, online user groups or news websites. 
For an $M$-dimensional multivariate Hawkes process, the intensity of the $m$-th dimension takes the following form:  
\begin{align}\label{eq:intensity}
\lambda_m(t) = \mu_m + \sum_{j: t_j<t} g_{m,m_j}(t-t_j),  
\end{align}
where $\mu_m$ is the background rate of dimension $m$ and $g_{m,m_j}(\cdot)\geq 0$ is the triggering kernel 
that captures the impact of an event in community $m_j$ on the occurrence of an event in community $m$.  
The typical choice for the triggering kernel is the exponential memory kernel, which is defined by  
$g_{m,m_j}(\Delta_j) = \alpha_{m,m_j} \exp{(-\beta_{m,m_j} \Delta_j)}$, 
where $\Delta_j$ represents the time interval $\Delta_j = t-t_j$,  
$\alpha_{m,m_j}$ quantifies the magnitude of the influence from community $m_j$ on the event occurrence in community $m$, 
and $\beta_{m,m_j}$ controls how quickly its effect decays in time (i.e., speed of the diffusion).  
Other candidates include power law kernel \cite{ogata1999seismicity}, Raleigh kernel \cite{wallinga2004different}, and log-normal distribution \cite{navaroli2015modeling}.  

The negative log-likelihood function of a multivariate Hawkes process over time interval $[0,T]$ is given by:
\begin{align}\label{eq:define_loglike}
\mathcal{L} &= \sum_{i=1}^I \log{\lambda_{m_i}(t_i)} - \sum_{m=1}^M \int_{0}^{T} \lambda_{m}(t) dt. 
\end{align}

\subsection{Problem Definition}
The key notations used in the paper are listed in Table \ref{tab:symbols} of Appendix \ref{sec:notation}.  
An event is represented by the pair $(t,m)$, where $t$ and $m$ denote time and community (e.g., country, news website) where the event happened, respectively.  
An event sequence is defined as the set of events $\mathcal{D} = \{(t_i,m_i)\}_{i=1}^I$ with $t_i\in[0,T)$, 
where $I$ denotes the number of events that have occurred up to time $T$.  

\textbf{Event Prediction Problem. }
Given the event sequence $\mathcal{D}$ in the observation time window $[0,T)$,  
we aim to leverage $\mathcal{D}$ to predict the number of events within any given time period; and the event times in the future time window $[T, T+\Delta T]$.

\section{Dynamic Hawkes Processes}
\begin{figure}[t]
  \includegraphics[width=0.63\linewidth]{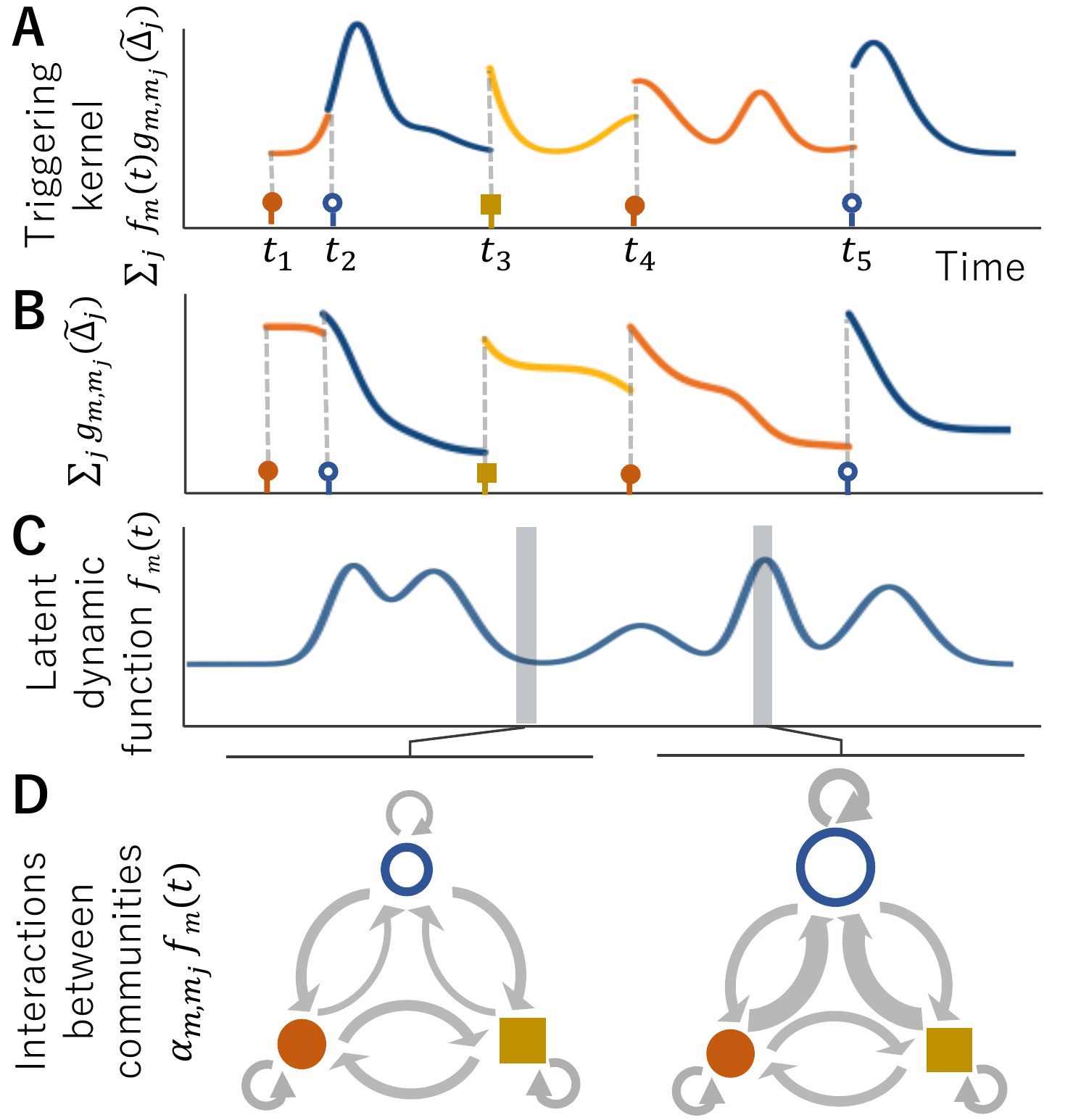}
\vspace{-1mm}
\caption{An illustration of DHP. Panel A represents individual events (i.e., input event sequence) and 
the dynamic interaction term for dimension \textcolor{blue(pigment)}{$\circ$}.  
The different dimensions (i.e., communities) are shown in different colors and markers. 
Panels B and C show the modified triggering kernel $\sum_{j}g_{m,m_j}(\tilde{\Delta}_j)$ and the latent dynamic function $f_m(t)$ for dimension \textcolor{blue(pigment)}{$\circ$}, respectively. 
Panel D depicts the interactions between communities at two different times.  
The size of each node represents the level of activity; the width of each arrow represents the level of interaction at each time. 
}\label{fig:overview} 
\vspace{-5.5mm}
\end{figure}

In this section, we present \textsf{DHP} (Dynamic Hawkes Process), a novel multivariate Hawkes process framework for event prediction; it can learn the time-evolution of the communities underlying the diffusion processes.
Figure~\ref{fig:overview} illustrates DHP. 
We design the triggering kernel of \textsf{DHP} intensity (panel A in Figure \ref{fig:overview}) as the product of two components: 
the triggering kernel with the input of time-rescaled events (panel B in Figure \ref{fig:overview}), 
which learns the decay influence from the past events; and 
the {\sl latent dynamics function} (panel C) to adjust the magnitude of the influence from the past events. 
The latent dynamics function describes the time-evolving states of the communities (indicated by dimensions).  
In the context of disease spread, the latent dynamics function represents the dynamics of people's awareness of the disease in each country.  
For information diffusion, it characterizes the temporal evolution of readers' interests in news websites. 
We elaborate on the formulation of \textsf{DHP} in $\S \ref{sec:formulation}$, 
followed by parameter learning ($\S$ \ref{sec:learning}). 
The prediction procedure is described in Appendix \ref{sec:pred}.  

\subsection{Model Formulation}\label{sec:formulation}
The proposed model specifies the intensity of Hawkes process for dimension $m$ as 
\begin{align}\label{eq:lamda}
\lambda_m(t) &= \mu_m + \sum_{j: t_j<t} g_{m,m_j}(\tilde{\Delta}_j) f_m(t), 
\end{align}
where $\mu_m$ is the background rate for the $m$-th dimension (i.e., community), $g_{m,m_j}(\cdot)$ is any chosen triggering kernel 
between dimension $m$ and dimension $m_j$    
such as exponential memory kernel or log-normal distribution,  
and $\tilde{\Delta}_j$ is the time-rescaled or transformed time interval between the current time $t$ and the time of $j$-th event $t_j$.   
$f_m(t)\geq 0$ represents the dynamics of the $m$-th community underlying the diffusion processes at time $t$, 
which controls the magnitude of diffusion. 
The transformed time interval $\tilde{\Delta}_j$ is defined by the integral of the latent dynamics function between $t_j$ and $t$ as follows: 
\begin{align}\label{eq:kernel}
\tilde{\Delta}_j = \int^{t}_{t_j} f_m(\tau) d\tau = F_m(t) - F_m(t_j), 
\end{align}
where $F_m(t)$ denotes the integral function of the continuous-time dynamics $f_m(t)$, that is $F_m(t)=\int_{0}^t f_m(\tau)d\tau$. 
The above formulation can be understood by considering an analogy drawn from the time-rescaling theorem \cite{brown2002time}.  
Intuitively, this transformation adjusts the influence of each event by stretching or shrinking time based on the value of the latent dynamics function $f_m(t)$.  
When $f_m(t)<1$, the interval times are lengthened so that event times are further separated. Likewise, when $f_m(t)>1$, the interval times are compressed so that events are drawn closer together. If $f_m(t)=1$ for all $t>0$, $\tilde{\Delta}_j=\int_{t_j}^t 1 d\tau = t-t_j$, and Equations \ref{eq:lamda} and \ref{eq:kernel} reduce to a simple multivariate Hawkes process (Equation \ref{eq:intensity}).  
This formulation assumes that the speed of diffusion varies according to the temporal dynamics of the target community $m$,  which is captured by $f_m(t)$.  
This assumption is realistic; for instance, disease spread is controlled by people's awareness of the disease in each country and their preventive behaviors.   
Information diffusion is largely influenced by the reader's interest in each news website. 
It is worth mentioning that the latent dynamic function can be easily extended to consider the dynamics of pairwise interactions between dimensions, 
by redefining the latent dynamics function as $f_{m,m_j}(t)$. 
The following discussion holds even under this extension. 

Our formulation allows considering the latent state of each dimension $m$ at current time $t$ as well as the influence from the past events.  
Also, it can capture the simultaneous changes in diffusion magnitude and speed,  
which is desirable for many applications (as discussed in the following paragraph). 
Most importantly, it enables us to compute the analytic integral of the intensity, which is required for evaluating the log-likelihood 
(further discussion can be found in $\S$ \ref{sec:learning}) and predicting the number of future events (See Appendix \ref{sec:pred}). 

{\bf Triggering Kernel.}  
The triggering kernel can have many forms.
For example, we can assume the exponential memory kernel for $g_{m,m_j}(\cdot)$, i.e.,  
{\small\begin{align}\label{eq:exp}
\lambda_m(t) &= \mu_m + \sum_{j: t_j<t} f_m(t) \alpha_{m,m_j}\exp{\left( -\beta_{m,m_j} \big(F_m(t)-F_m(t_j)\big) \right)}, 
\end{align}}
\hspace{-1mm}where $\alpha_{m,m_j}$ encompasses the magnitude of the static interaction between the $m$-th and $m_j$-th dimension; and   
$\beta_{m,m_j}$ weights the decay of the influence over time.  
Notice that, the above formulation relies on the implicit assumption that the magnitude and speed of diffusion are related through the latent dynamics function $f_m(t)$,  which controls the magnitude of diffusion; its integral $F_m(t)$ governs the speed of diffusion.  
For example, when $f_m(t)=2$ for every $t$, 
the second term of Equation \ref{eq:exp} is $\underline{2\alpha}\exp{(-\underline{2\beta}\Delta_j)}$, where $\Delta_j=t-t_j$.  
When $f_m(t)=0.5$ for every $t$, it is $\underline{0.5\alpha}\exp{(-\underline{0.5\beta}\Delta_j)}$. 
This assumption is reasonable since the magnitude and speed of diffusion vary simultaneously in many cases.  
Taking disease transmission as an example, active prevention measures can both reduce the magnitude and the speed of the infection.
The influence on the magnitude of diffusion from the latent dynamics function is tuned by $\alpha$, and  
the influence on the speed of diffusion from its integral is tuned by $\beta$.  

{\bf Latent dynamics function.} The design of $f_m(\cdot)$ is flexible to so any non-negative function can be used. 
Inspired by \cite{omi2019fully}, we utilize and extend a monotonic neural network \cite{sill1998monotonic,chilinski2020neural} that learns a strictly monotonic function to design the latent dynamics function. 
Concretely, we model the integral function $F_m(t)$ using the monotonic neural network. 
This guarantees that its derivative (i.e., the latent dynamics function $f_m(t)$) is strictly non-negative,  
so intensity $\lambda_m(t)$ results in a non-negative function. 
In describing the integral function we propose to further enhance the expressiveness of the monotonic neural network by using a mixture of monotonic neural networks. 
Formally, \begin{align}\label{eq:mixture}
F_m(t) = \sum_{c=1}^C \pi_c \Phi_m^c(t) + b_0 t,  
\end{align}
where $C$ is the number of mixture components, $\Phi_m^c(\cdot)$ is the $c$-th monotonic neural network, $\pi_c$ is the mixture weight of the $c$-th component,  
and $b_0$ is a bias parameter for the output layer. 
To preserve monotonicity of the integral of the latent dynamics function $F_m(t)$, we impose non-negative constraints on the mixture weights $\{\pi_1,...,\pi_C\}$  
and parameter $b_0$.  
For each dimension, we construct $L$ fully connected neural layers with monotonic activation functions.  
Whenever the context is clear, we simplify notation $\Phi_m^c(\cdot)$ to $\Phi(\cdot)$.
At each layer $l\in\{1,2,...,L\}$ of the monotonic neural network, the hidden-state vector ${\bf h}^{(l)}$ is given by 
\begin{align}
{\bf h}^{(l)} = \sigma({\bf W}^{(l)} {\bf h}^{(l-1)} + {\bf b}^{(l)}), \end{align}
where ${\bf W}^{(l)}$ and ${\bf b}^{(l)}$ are parameter matrix and vector to learn for $l$-th layer, respectively.  
The input of the first layer is time $t$: ${\bf h}^{(0)} = t$.  
$\sigma(\cdot)$ is a monotonic non-linear function. 
Following the previous work \cite{sill1998monotonic,chilinski2020neural}, we use {\sl tanh} activation for hidden layers and {\sl softplus} for the last layer.   
The output of the monotonic neural network is  
$\Phi(t) = {\bf B} {\bf h}^{(L)}$, 
where ${\bf B}$ is a learnable weight matrix. 
The weight parameter matrices ${\bf W}^{(l)}$ and ${\bf B}$ are imposed to be non-negative. 
The latent dynamics function $f_m(t)$, which is the derivative of the monotonic neural network $F_m(t)$, takes the following form, 
\begin{align}
f_m(t) = \sum_{c} \pi_c \phi_m^c(t) + b_0,  
\end{align}
where $\phi_m^c(t)$ is the gradient of the monotonic neural network $\Phi_m^c(t)$ with respect to time $t$, namely $\phi_m^c(t)=d\Phi^c_m(t)/dt$. 
The gradient $\phi_m^c(t)$ can be obtained by applying the automatic differentiation implemented in deep learning frameworks such as TensorFlow\cite{abadi2016tensorflow}.  
As we place no restriction on the parametric forms of the community dynamics underlying the diffusion processes,  
our model can fit various complex dynamics of each community's state. 

This design choice enables us to automatically learn unknown complex dynamics of the communities' states behind the diffusion processes, 
while at the same time allowing us to compute the exact log-likelihood for training as described in $\S$ \ref{sec:learning}.

\subsection{Parameter Learning}\label{sec:learning}
Given the history of events up to but not including $T$, $\mathcal{D}=\{(t_i,m_i)\}_{i=1}^I$, 
we learn all the parameters of \textsf{DHP} by minimizing the negative log-likelihood of the observed event sequences. 
Specifically, we simultaneously estimate the neural network weights and  
the kernel parameters $\{{\boldsymbol \mu},{\bf A},{\boldsymbol \beta}\}$: the background rates ${\boldsymbol \mu}=\{\mu_1,..,\mu_M\}$, 
the matrix of interactions ${\bf A}=(\alpha_{i,j})\in\mathbb{R}^{M\times M}$, and the decay rates ${\boldsymbol \beta}=\{\beta_1,..,\beta_M\}$. 
The negative log-likelihood function of \textsf{DHP} over  time interval of $[0,T)$ is 
obtained by substituting Equation \ref{eq:lamda} into Equation \ref{eq:define_loglike}: 
{\small\begin{align}
\mathcal{L} &= \sum_{i=1}^{I} \bigg[ \log{\Big( \mu_{m_i} + \sum_{j: t_j<t_i} g_{m_i,m_j} \big( F_{m_i}(t_i) - F_{m_i}(t_j) \big) f_{m_i}(t_i) \Big)} 
\\ \nonumber 
&- \sum_{m=1}^M \Big( \mu_{m}(t_i - t_{i-1}) + 
\underbrace{ \int_{t_{i-1}}^{t_i} \sum_{j: t_j<t} g_{m,m_j} \big( F_m(t) - F_m(t_j) \big) f_m(t) dt }_{\Lambda_i} \Big) \bigg]. 
\end{align}}
The problem here is to obtain integral $\Lambda_i$ in the last term, which reduces to
\begin{align}\label{eq:Lamda_i}
\Lambda_i &= \sum_{j: t_j<t_{i-1}} \int_{t_{i-1}}^{t_i} g_{m,m_j} \Big( \underbrace{F_m(t) - F_m(t_j)}_{\textrm{inner part}} \Big) f_m(t) dt.  
\end{align}
Notice that the integrand of the above integral can be regarded as the product of composite function $g_{m,m_j}(F_m(\cdot))$ and the derivative $f_m(t)$ of the ``inner'' part of that composite function.    
Hence, we can solve the integral of Equation \ref{eq:Lamda_i} in closed form by applying $u$ substitution (also called ``the reverse chain rule'') technique. Making the substitution $x=F_m(t)-F_m(t_j)$ gives $dx=f_m(t)dt$. 
Changing variables from $t$ to $x$, the integral of Equation \ref{eq:Lamda_i} becomes, \begin{align}\label{eq:deform}
&\int_{t_{i-1}}^{t_i} g_{m,m_j} \Big( F_m(t) - F_m(t_j) \Big) f_m(t) dt \\ \nonumber
&= \int_{F(t_{i-1})-F(t_j)}^{F(t_i)-F(t_j)} g_{m,m_j}(x) dx 
 = \left[ G_{m,m_j}(x) \right]_{F(t_{i-1})-F(t_j)}^{F(t_i)-F(t_j)} \\ \nonumber
&= G_{m,m_j}\left( F(t_i)-F(t_j) \right) - G_{m,m_j} \left( F(t_{i-1})-F(t_j) \right) 
    \end{align}
where $G_{m,m_j}(t)$ denotes the integral of $g_{m,m_j}(t)$.  
This can be computed analytically for many common kernels (See Appendix \ref{sec:kernel}). 

Given the exact log-likelihood, we back-propagate the gradients of the loss function $\mathcal{L}$.  
In the experiment, we employ mini-batch optimization.

\section{Experiments}
We start by setting up the qualitative and quantitative experiments, and then report their results. 
{\small\begin{table*}[!htb]
\caption{Statistics of Datasets used in this paper. }
\vspace{-4mm}
\begin{tabular}{lcccc} \toprule
        & Source & Time span & \# Events & Communities \\ \midrule
Reddit  & Reddit\footnoteref{Reddit API} & 1 Mar - 31 Aug, 2020 & 23,059 & 25 subreddits \\ 
News    & GDELT\footnoteref{GDELT} & 20 Jan - 24 Mar, 2020 & 19,541 & 40 news websites \\ 
Protest & ACLED\footnoteref{ACLED} & 1 Mar - 21 Nov, 2020 & 22,313 & 35 countries \\
Crime   & Chicago Data Portal\footnoteref{Chicago Portal} & 1 Mar - 19 Dec, 2020 & 29,318 & 13 community areas \\
\bottomrule 
\end{tabular}\label{tab:stats}
\end{table*}}
\subsection{Datasets} 
We used four real-world event datasets from different domains.  
\begin{itemize}
\item {\bf Reddit}:  
We crawled the official Reddit API\footnote{\label{Reddit API}Reddit API. http://www.reddit.com/dev/api. Accessed on December 10, 2020.} 
to gather timestamped hyperlinks between Reddit communities (i.e., subreddits) over 6 months from March 1 to August 31, 2020.
Following the work of \cite{nickel2020learning}, 
we use a list of hyperlinks to each target subreddit as a separate sequence and consider target subreddits as communities (i.e., dimensions). 
\item {\bf News}: News dataset, which is provided by GDELT project \cite{leetaru2013gdelt} through its API\footnote{\label{GDELT}Global Dataset of Events, Location,
and Tone (GDELT). http://gdeltproject.org. Accessed on December 23, 2020. }, 
consists of roughly 20,000 news articles related to COVID-19 dated from January 20 to March 24, 2020.
We filtered out 40 news websites and used them as communities. 
\item {\bf Protest}: Protest dataset, which was gathered by ACLED\footnote{\label{ACLED}Armed Conflict Location and Event Dataset (ACLED). https://www.acleddata.com. Accessed on December 10, 2020.}, contains over 20,000 demonstration events in 35 countries during 9 months from March 1 to November 21, 2020. 
\item {\bf Crime}: Crime dataset is publicly available from the City of Chicago Data Portal\footnote{\label{Chicago Portal}Chicago Data Portal. https://data.cityofchicago.org/. Accessed on December 30, 2020.}; it 
includes about 30,000 reported crimes from 13 community areas of Chicago from 1 March to 19 December 2020. 
We treat community areas as communities. 
\end{itemize}
All the datasets are publicly available. 
The statistics of these datasets are given in Table \ref{tab:stats}. 
The procedure of data preprocessing is provided in Appendix \ref{sec:datasets}.

\subsection{Comparison methods}
We compare \textsf{DHP} against five widely used point process methods that incorporate the influence of the past events: 
\begin{itemize}
\item \textsf{HPP} (Homogeneous Poisson Process): It is the simplest point process where the intensity 
is assumed to be constant over time. 
\item \textsf{RPP} (Reinforced Poisson Processes) \cite{shen2014modeling,pemantle2007survey}: 
\textsf{RPP} accounts for the aging effect and the cumulative count of past events.  
\item \textsf{SelfCorrecting} (Self-correcting Point Process) \cite{isham1979self}: 
Its intensity is assumed to increase linearly over time and this tendency is corrected by the historical events.  
\item \textsf{Hawkes} (Hawkes Process): Its intensity is parameterized by Equation \ref{eq:intensity}, 
which explicitly models the influence of the past events by using the static triggering kernel.  
\item \textsf{RMTPP} (Recurrent Marked Temporal Point Process) \cite{du2016recurrent}: It employs RNN to encode the non-linear effects of past events. 
The event sequences are first embedded by RNN and then used as the input of the intensity. 
\end{itemize}
See Appendix \ref{sec:baselines} for details of the comparison methods.

\subsection{Experimental Settings}
For the experiments, we divided each dataset into train, validation, and test sets by chronological order with the ratios of 70\%, 10\%, and 20\%. 
The model parameters were trained using the ADAM optimizer \cite{kingma2014adam}.  
We tuned all the models using early stopping based on the log-likelihood performance on the validation set 
with a maximum of 100 epochs for the Reddit and News datasets and 30 epochs for the Protest and Crime datasets. 
Batch size is set to 128. 
The hyperparameters of each model are optimized via grid search.  
For the neural networks-based models (i.e., \textsf{RMTPP} and \textsf{DHP}), we choose the number of layers from $\{1,2,3,4,5\}$.  
For Hawkes process methods (i.e., \textsf{Hawkes} and \textsf{DHP}), the kernel function is selected from three commonly used kernels: 
exponential memory, power-law, and Raleigh kernels.  
These are mathematically defined in Appendix \ref{sec:kernel}. 
For \textsf{DHP}, we search on the number of mixtures $C$ over $\{1,2,3,4,5\}$. 
The chosen hyperparameters are presented in Appendix \ref{sec:impl_detail}. 

\subsection{Evaluation Metrics}
Our experiments use the following two metrics in evaluating all models. 
For both metrics, lower values indicate better performance.
\begin{itemize}
\item {\bf NLL} (Negative Log-Likelihood) is used to assess the likelihood of the occurrence of the events over the test period; it is calculated as 
$\sum_{i=I}^{I+n} \left[- \log{\lambda_{m_i}(t_i)} + \sum_{m=1}^M \int_{t_{i-1}}^{t_i} \lambda_{m}(t) dt \right]$, 
where $n$ is the number of events in the test period. 
\item {\bf MAPE} (Mean Absolute Percentage Error) evaluates the discrepancies between the predicted number of events in small time intervals and the ground truth.  
We first split the test time period $[T,T+\Delta T]$ into $S$ successive small time intervals using 15-minute periods. 
For each time interval $(t_s,t_{s+1}]$ and each dimension $m$, given the history of events up to $t_s$, we predict the number of events in $(t_s,t_{s+1}]$,  
$\hat{N}^m((t_s,t_{s+1}])$, described in Equation \ref{eq:pred} of Appendix \ref{sec:pred}.  
Then, we measure the average normalized difference between the predicted and observed number of events across all time intervals as follows: 
{\small\begin{align}
\text{MAPE}=\frac{1}{M}\sum_{m=1}^M 
\frac{\big|\sum_{s=1}^S\hat{N}^m\big( (t_s,t_{s+1}] \big) - \sum_{s=1}^S N^m\big( (t_s,t_{s+1}] \big)\big| }{\sum_{s=1}^S N^m\big( (t_s,t_{s+1}] \big)}, 
\end{align}}
\hspace{-1mm}where $\hat{N}^m\left((t_{s+1},t_s]\right)$ is the predicted number of events in the small time interval $(t_{s+1},t_s]$ and $N^m(\cdot)$ is the ground truth at the $s$-th time interval and $m$-th dimension.   
\end{itemize}

{\small\begin{table*}
\centering\caption{
Negative log-likelihood (NLL) and Mean Absolute Percentage Error (MAPE) with standard deviation (in the bracket). Lower is better. 
The best performance is in bold. 
Our proposal, \textsf{DHP}, outperforms five existing methods. }
\begin{tabular}{lcccccccc} \toprule
      & \multicolumn{2}{c}{Reddit} & \multicolumn{2}{c}{News} & \multicolumn{2}{c}{Protest} & \multicolumn{2}{c}{Crime} \\
      & NLL & MAPE & NLL & MAPE & NLL & MAPE & NLL & MAPE \\ \midrule
\textsf{ HPP            } & -5.637 & 0.553 (0.204) & -5.710 & 0.600 (0.044) & -5.753 & 0.345 (0.060) & -6.795 & 0.144 (0.014) \\
\textsf{ Hawkes         } & -5.696 & 0.458 (0.107) & -6.167 & 0.471 (0.085) & -6.260 & 0.415 (0.371) & -6.799 & 0.179 (0.016) \\
\textsf{ RPP            } & -5.568 & 0.595 (0.259) & -6.150 & 0.481 (0.522) & -5.643 & 0.581 (0.759) & -6.781 & 0.175 (0.021) \\
\textsf{ SelfCorrecting } & -5.662 & 0.475 (0.158) & -5.973 & 0.452 (0.059) & -5.750 & 0.524 (0.674) & -6.803 & 0.123 (0.005) \\
\textsf{ RMTPP          } & - & 0.311 (0.061) & - & 0.446 (0.125) & - & 0.639 (1.337) & - & 0.302 (0.010) \\ \midrule
\textsf{ Proposed       } & {\bf -6.447} & {\bf 0.305} (0.045) & {\bf -6.301} & {\bf 0.442} (0.039) & {\bf -6.914} & {\bf 0.318} (0.049) & {\bf -6.983} & {\bf 0.117} (0.008) \\
\bottomrule 
\end{tabular}\label{tab:accuracy}
\end{table*}}

\subsection{Performance Evaluation}
In this section, we first compare \textsf{DHP} with existing methods on event prediction. 
Table \ref{tab:accuracy} presents the negative log-likelihood (NLL) of the test data and Mean Absolute Percentage Error (MAPE) 
for different methods on the real-world event datasets.  
In this table, we omit the result of \textsf{RMTPP} since its log-likelihood function differs from those used in the other methods,  
(it is defined for the whole event sequence from all communities, not for the separate sequences of the individual communities,  
which precludes fair comparison). 
As shown in the table, our proposal, \textsf{DHP}, outperforms the four existing methods across all the datasets in terms of NLL. 
\textsf{HPP} has the worst NLL in most cases since it does not explore the temporal variation of the event occurrences.
\textsf{RPP} and \textsf{SelfCorrecting} cannot achieve good results as they 
encode strong assumptions on the functional forms of the intensity, which limits the expressivity of the model. 
\textsf{Hawkes} surpasses \textsf{HPP}, \textsf{RPP}, and \textsf{SelfCorrecting}, which explicitly models the dependencies between past and current events.  
However, it still falls short for modeling the dynamic changes of the community states in the diffusion process. 
Our \textsf{DHP} achieves even better NLL than \textsf{Hawkes}.  
This verifies that incorporating latent community dynamics is essential for event prediction and 
that \textsf{DHP} can learn effective representations of the time-evolving dynamics of community states. 

\textsf{DHP} achieves the best MAPE for all datasets.  
\textsf{RMTPP} performs the second best in terms of MAPE for Reddit and News datasets, 
which is probably because \textsf{RMTPP} exploits the power of RNN for learning non-linear dependencies between events. 
But \textsf{RMTPP} performs poorly for Protest and Crime datasets since it cannot capture changes in the event occurrences 
due to the temporal evolution of communities' states,  
e.g., a large reduction in protest events due to the COVID-19. 
\textsf{DHP} outperforms all other methods across the datasets on the two metrics.  
The above result reveals the effectiveness of encoding the community state dynamics governing the diffusion process for event prediction. 
It also suggests that the assumption of \textsf{DHP}, i.e., the magnitude and speed of diffusion are related, holds for real diffusion processes.

\subsection{Sensitivity Study}\label{sec:sensitivity}
In this section, we analyze the impacts of hyperparameters or experimental settings. 
We report the prediction performance of \textsf{DHP} under different settings for the four datasets. 

\begin{figure}[t]
\minipage{0.25\textwidth}
  \centering\includegraphics[width=0.85\linewidth]{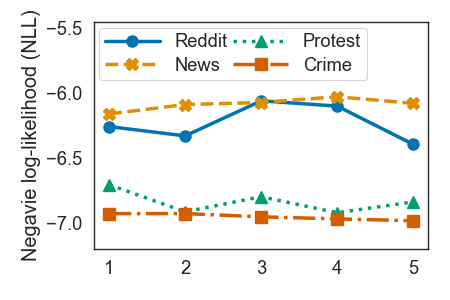}\vspace{-2.5mm}
  \subcaption{Number of mixtures}\label{fig:mixture}
\endminipage\minipage{0.25\textwidth}
  \centering\includegraphics[width=0.85\linewidth]{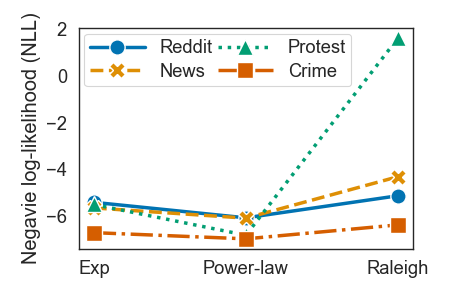}\vspace{-2.5mm}
  \subcaption{Triggering kernel}\label{fig:kernel}
\endminipage\hfill
\minipage{0.25\textwidth}
  \centering\includegraphics[width=0.85\linewidth]{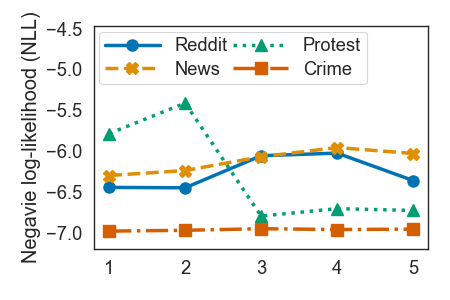}\vspace{-2.5mm}
  \subcaption{Number of layers}\label{fig:layer}
\endminipage
\vspace{-2mm}
\caption{Sensitivity Study: NLL performance of \textsf{DynamicHawkes} on different settings for four datasets. }
\vspace{-5mm}
\end{figure}

{\bf Number of mixtures. } 
We examine how the number of mixture components, $C$, determines the prediction performance of \textsf{DHP}.  
Figure \ref{fig:mixture} shows the negative log-likelihood (NLL) on the test data with respect to different numbers of mixtures $\{1,2,3,4,5\}$. 
In this experiment, we fixed the number of layers as 3 and used the power-law kernel.  
The NLL performance tends to be stable for all the datasets. 
It slightly increases as the number of mixture components becomes larger for Protest and Crime dataset. 
The results indicate that increasing the number of mixtures can improve the expressiveness of the model. 

{\bf Kernel functions. } 
We investigate the effect of three kernel functions: exponential kernel, power-law kernel, and Raleigh kernel, 
where the number of mixtures and the number of layers are set to 3.  
As shown in Figure \ref{fig:kernel}, the power-law kernel yields the best performance on all datasets. 

{\bf Number of layers. } 
Figure \ref{fig:layer} evaluates the sensitivity of our neural network $\Phi_m^c(t)$ to the number of layers $L\in\{1,2,3,4,5\}$  
by fixing the number of mixtures as 3 and using the power-law kernel. 
We observe that \textsf{DHP} yields better NLL results for the Protest dataset with larger numbers of layers.   
For the other three datasets, it has little effect on the performance. 

In general, \textsf{DHP} shows stable and robust prediction performance across different settings. 

\begin{figure}[t]
    \hspace{5mm}\includegraphics[width=0.92\linewidth]{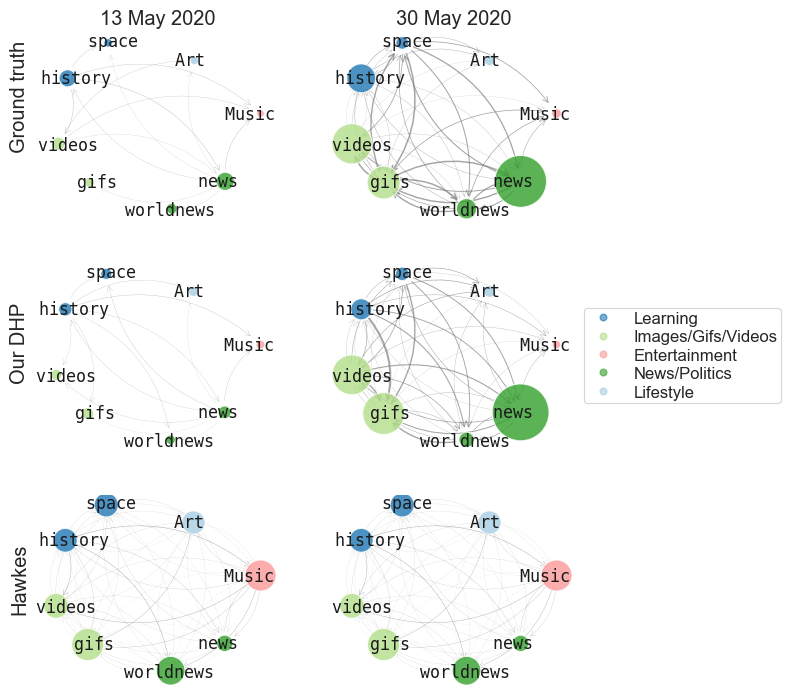}
    \vspace{-1mm}
    \caption{Visual comparison of predicted interactions among reddit communities (i.e., subreddits) from Reddit dataset. 
The bottom two rows are the predicted results of our \textsf{DHP} and \textsf{Hawkes}. 
The last row is the ground truth. 
Columns correspond to times indicated by the label on the top. Nodes represent subreddits. Their colors indicate their categories. 
For \textsf{DHP} (middle), the size of the $m$-th node is proportional to $\sum_{m'} \alpha_{m,m'} f_m(t)$ and 
the width of edge between nodes $m$ and $m'$ is proportional to $\alpha_{m,m'} f_m(t)$.  
    }\label{fig:vis_network}
\vspace{-6mm}
\end{figure}

\subsection{Case Studies}
In order to further verify the capability of \textsf{DHP}, we analyze the temporal dynamics of the community states behind the diffusion process learned by \textsf{DHP} from each dataset.  

\begin{figure}[t]
\minipage{\linewidth}
  \centering\includegraphics[width=0.82\linewidth]{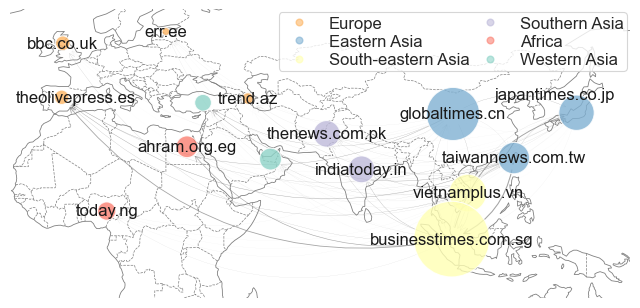}
  \subcaption{February 24, 2020. }\label{fig:news_newtwork_sampled_0}
\endminipage\hfill
\minipage{\linewidth}
  \centering\includegraphics[width=0.82\linewidth]{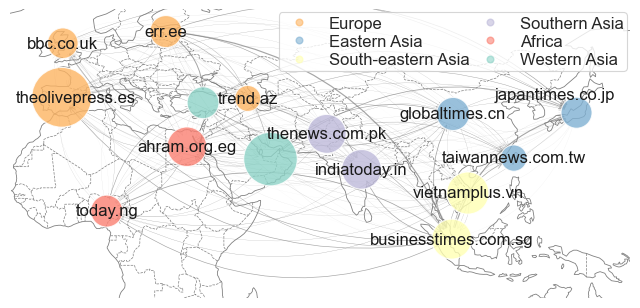}
  \subcaption{March 15, 2020. }\label{fig:news_newtwork_sampled_1}
\vspace{-2mm}
\endminipage\hfill
\minipage{\linewidth}
\endminipage\hfill
  \caption{
Inferred interactions among 15 major news websites from different countries by \textsf{DHP} on News dataset at two different time points.  
Nodes refer to domain names of news websites. 
We used the top-level domain to specify the country in which each news website is based.  
Notes are colored by regions.  
The size of the $m$-th node is given by $\sum_{m'} \alpha_{m,m'} f_m(t)$ and 
the edge width between nodes $m$ and $m'$ by $\alpha_{m,m'} f_m(t)$.  
  }\label{fig:news_network_sampled}
\vspace{-2mm}
\end{figure}
\begin{figure}[t]
  \centering\includegraphics[width=0.91\linewidth]{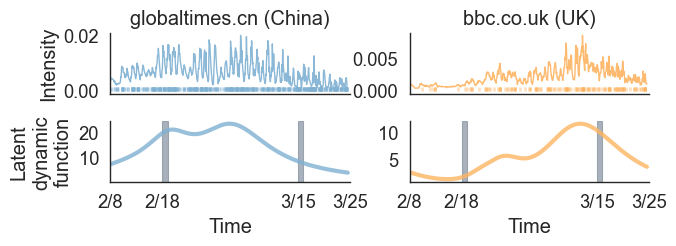} \vspace{-4mm}
  \caption{Learned intensity and latent dynamic function for two news websites in China and UK from February 14 to March 25, 2020. }\label{fig:news_ft_sampled}
\vspace{-5mm}
\end{figure}
Figure \ref{fig:vis_network} visualizes the interactions between selected 8 Reddit communities (i.e., subreddits) learned from Reddit dataset.  
In each row, we compare the estimated interactions between subreddits by \textsf{DHP} (middle) and \textsf{Hawkes} (bottom), and the ground truth (top).  
For the ground truth, node size corresponds to the aggregated number of hyperlinks for each ``target'' community in the default 5-day interval; 
the weight of each edge represents the number of hyperlinks between source community $m'$ and target community $m$.  
For \textsf{DHP}, node size is proportional to $\sum_{m'} \alpha_{m,m'} f_m(t)$; edge width is $\alpha_{m,m'} f_m(t)$. 
For \textsf{Hawkes}, node size is $\sum_{m'}\alpha_{m,m'}$; edge width is $\alpha_{m,m'}$. 
Note that \textsf{Hawkes} produces the same results across times since it assumes the triggering kernel is static over time.  
We can see that the interactions learned by \textsf{DHP} are more consistent with the true evolution of the interactions 
between online user communities compared to \textsf{Hawkes}. 

The top panel of Figure \ref{fig:reddit_five_sample} shows the intensity $\lambda_m(t)$ and estimated dynamics function $f_m(t)$ learned for Reddit dataset  
along with the observed event sequences for the two subreddits. 
The latent dynamics function increases up to the end of May, rapidly for \texttt{news} and slowly for \texttt{space}. 
This is probably due to the COVID-19 lockdown. 
These results demonstrate that our \textsf{DHP} learns a reasonable representation of the latent temporal dynamics of the online communities. 

Figure \ref{fig:news_network_sampled} shows inferred interactions among news websites from 15 countries learned for News dataset. 
In these figures, the node size denotes the value of the latent dynamics function $\sum_{m'} \alpha_{m,m'}f_m(t)$ for each news website; 
the edge width denotes the strength of interactions between them $\alpha_{m,m_j}f_m(t)$. 
East Asian and South-East Asian countries (denoted by blue and yellow) rise to their peaks around late February (See Figure \ref{fig:news_newtwork_sampled_0}) 
and then decrease until mid-March (Figure \ref{fig:news_newtwork_sampled_1}), 
while the other countries are peaked around or after March 15 (Figure \ref{fig:news_newtwork_sampled_1}), not in February (Figure \ref{fig:news_newtwork_sampled_0}). 
We can also see in Figure \ref{fig:news_ft_sampled} that the dynamic function peaks around mid-February for China (left), 
followed by the United Kingdom with the peak of mid-March (right). 
These trends are synchronized to the growth of the pandemic in each country.  
East Asian and South-East Asian countries experienced their first peak in COVID-19 cases ahead of the other countries,  
which would trigger the people's early interest on COVID-19 related topics and accelerate the spread of COVID-19 related news early on.  
This confirms that our proposal, \textsf{DHP}, well reproduces the complex evolution in news website activities. 

\begin{figure}[t]
  \includegraphics[width=0.88\linewidth]{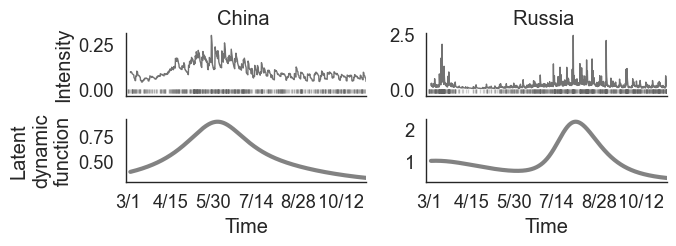}\vspace{-5mm}
  \caption{Intensity and latent dynamic function learned by \textsf{DHP} on Protest dataset for two countries. }\label{fig:protest_ft_sampled}
\vspace{-5mm}
\end{figure}

Figure \ref{fig:protest_ft_sampled} shows the intensity and latent dynamic function learned from the Protest dataset.  
According to a previous study\footnote{https://acleddata.com/2020/09/03/demonstrations-political-violence-in-america-new-data-for-summer-2020/}, 
in contrast to the online events, 
the pandemic initially leads to a reduction in protest events and the trend was corrected after several weeks. \textsc{DHP} well characterizes this trend. 
In China (left), the dynamic function decreased following the onset of the coronavirus around the beginning of March  
and returned to a moderate level by mid-June.  
For Russia, it declined gradually from March until the beginning of July, where the first peak of the pandemic occured around May 11.  
In conclusion, \textsf{DHP} uncovers the latent community dynamics underlying the diffusion processes, 
and so provides meaningful insights about the diffusion mechanism.

\section{Conclusion and Future work}
Modeling and predicting diffusion processes are important tasks in many applications. 
We presented a novel Hawkes process framework, \textsf{DHP} (Dynamic Hawkes Process), that can learn the temporal dynamics of the community states underlying diffusion processes. 
The proposed \textsf{DHP} allows for the automatic discovery of the community state dynamics underlying the diffusion processes as well as offering tractable learning.  
By conducting extensive experiments on four real event datasets, 
we demonstrate that \textsf{DHP} provides better performance for modeling and predicting diffusion processes than several existing methods. 
 
For future work, we plan to explore the following two directions.  
First, \textsf{DHP} can be extended to capture the pairwise dynamics of the interactions among communities  
by introducing the latent dynamics function for pairs of communities.
We will extend \textsf{DHP} to this case and conduct experiments to evaluate the performance of the extended \textsf{DHP} 
in capturing the time-evolving dynamics of the pairwise interactions between communities.  
Secondly, \textsf{DHP} is built on the assumption that the magnitude and speed of the diffusion are related to each other, 
which may limit the flexibility of the model. 
We will explore how to modify \textsf{DHP} to ease this assumption.

\bibliographystyle{ACM-Reference-Format}
\bibliography{KDD21} 

\clearpage
\appendix
\section*{Supplemental Material}
\section{Notation reference}\label{sec:notation}
For readers' convenience, we list the important notations used throughout the paper in Table \ref{tab:symbols}.  
{\small\begin{table}
\centering\caption{Table of symbols. }
\begin{tabular}{c|l} \toprule
Symbol & Definition \\ \midrule
$\mathcal{H}(t)$ & event sequence up to $t$ \\
$N(t)$ & total number of events up to $t$ \\
$N^m(t)$ & number of events of dimension $m$ up to $t$ \\
$\lambda_{m}(t)$ & intensity function for dimension $m$ \\
$\mu_m$ & background rate for dimension $m$ \\
$g_{m,m'}(t)$ & triggering kernel between dimension $m$ and dimension $m'$ \\ 
$\alpha_{m,m'}$ & interactions between dimension $m$ and demension $m'$ \\
$f_m(t)$ & dynamic function for dimension $m$ \\
$F_m(t)$ & integral of dynamic function for dimension $m$ \\
$\Phi_m^c(t)$ & neural network function of dimension $m$ for component $c$ \\ 
$M$ & number of dimensions \\
$C$ & number of mixture components \\
$L$ & number of layers of neural network \\
$S$ & number of time intervals in test time period \\ \bottomrule 
\end{tabular}\label{tab:symbols}
\end{table}}

\section{Kernel integral}\label{sec:kernel}
In our experiment, we used three types of triggering kernel: Exponential, Power-law and Raleigh.  
Table \ref{tab:kernels} presents their equations and integrals, where $\alpha$ and $\beta$ are parameters of the triggering kernel.
$p$ is the scaling exponent of the power-law ($p>1$) and we fix $p=2$ in the experiment.   

\begin{table}
\centering\caption{Integral for some common kernels. }
\begin{tabular}{ccc} \toprule
Type & Equation $g(t)$ & Integral $G(t)$ \\ \midrule
Exponential (EXP) & $\alpha \exp{(-\beta t)}$ & $-\frac{\alpha}{\beta} \exp{(-\beta t)}$ \\
Power-law (PWL) & $\frac{ \alpha\beta }{ (\alpha+\beta t)^{(p+1)} }$ & $-\frac{\alpha}{p(\alpha+\beta t)^p}$ \\
Raleigh (RAY) & $\alpha  t \exp{(-\beta  t^2)}$ & $-\frac{\alpha}{2\beta} \exp{(-\beta t^2)}$ \\
\bottomrule 
\centering
\end{tabular}\label{tab:kernels}
\vspace{-4mm}
\end{table}

\section{Preidiction}\label{sec:pred}
For each time interval $(t_s,t_{s+1}]$ and each dimension $m$, 
given the history of events up to time $t_s$, we calculate the expected number of events in $(t_s,t_{s+1}]$ by $\int_{t_s}^{t_{s+1}} \lambda_m(\tau)\tau$. 
As discussed in Section \ref{sec:learning}, this integral takes analytic form for the proposed method. 
Similarly to Equation \ref{eq:deform}, we obtain  
{\small\begin{align}\label{eq:pred}
&\hat{N}^m((t_s,t_{s+1}]) = \int_{t_s}^{t_{s+1}} \lambda_m(\tau)\tau = \mu_m(t_{s+1}-t_s) \\ \nonumber
&+ \sum_{j:t_j<t_s} G_{m,m_j}\left( F(t_{s+1})-F(t_j) \right) - G_{m,m_j} \left( F(t_s)-F(t_j) \right),  
\end{align}}
\hspace{-1mm}where $\hat{N}^m\left((t_{s+1},t_s]\right)$ is the predicted number of events in the given time interval $(t_{s+1},t_s]$ for dimension $m$. 

\section{Experiment}
\subsection{Datasets}\label{sec:datasets}
\begin{itemize}
\item {\bf Reddit}: 
The data collection procedure followed the one used in \cite{kumar2018community}.  
During crawling we selected the 25 most popular subreddits, and retrieved hyperlinks among those subreddits: 
we identified and recorded posts in one source subreddit that contain links to different target subreddits. 
This process finally yielded a total of roughly 23,000 posts, each of which had submission time, source subreddit, and target subreddit.  
We treated a list of hyperlinks to each target subreddit as a separate sequence and considered target subreddits as communities (i.e., dimensions). 
The source subreddit was not used for training but for qualitative evaluation (Figure \ref{fig:vis_network}).  
\item {\bf News}: 
The original dataset contains over a million of news articles related to COVID-19. 
Each piece of news had a timestamp and a URL. 
We extracted the domain of news websites from a URL and obtained more than 1,000 unique domains. 
We filtered out 40 country-specific domains and used them as communities. 
The granularity of time is one second. 
\item {\bf Protest}: 
We sampled 35 popular countries and retrieved events from those countries.  
Each event was associated with two attributes: timestamp and country. 
The dataset was recorded at minute level.  
\item {\bf Crime}: 
Each event recorded the time and community area where a crime happened. 
The time granularity is one minute. 
\end{itemize}

\subsection{Implementation details}\label{sec:impl_detail}
All code was implemented using Python 3.9 and Keras \cite{chollet2015keras} with a TensorFlow backend \cite{abadi2016tensorflow}. We conducted all experiments on a machine with four 2.8GHz Intel Cores and 16GB memory.  

The model parameters were trained using the ADAM optimizer \cite{kingma2014adam} 
with $\beta_1=0.9$, $\beta_2=0.999$ and a learning rate of 0.002. 
For the neural networks-based models (i.e., \textsf{RMTPP} and \textsf{DHP}), the number of hidden units in each layer is fixed as 8.  
In our experiment, the number of mixtures is set to 3 for Reddit, News and Protest datasets, and to 5 for Crime dataset. 
In all experiments, we used the power-law kernel. 
The number of layers is set to 2 for Reddit and Protest datasets, 1 for News dataset, 3 for Crime dataset, respectively. 

\subsection{Comparison methods}\label{sec:baselines}
\begin{itemize}
\item \textsf{HPP} (Homogeneous Poisson Process): The simplest point process whose intensity is constant over time: $\lambda_m(t)=\lambda_m$.  
\item \textsf{RPP} (Reinforced Poisson Processes) \cite{shen2014modeling,pemantle2007survey}: 
For each dimension $m$, the intensity of \textsf{RPP} is characterized by  
\begin{align}
\lambda_m(t) = \gamma_m(t) N^m(t), \end{align}
where $\gamma_m(t)$ is the relaxation function that characterizes the aging effect, and  
$N^m(t)$ is the number of events of dimension $m$ that have occurred up to $t$.  
Following the prior work \cite{shen2014modeling}, we define $\gamma_m(t)$ by the following relaxation log-normal function:  
\begin{align}
\gamma_m(t) = \frac{ \exp{( -(\log{t}-\alpha_m)^2/2\beta_m^2 )} }{ \sqrt{2\pi}\beta_m t },  
\end{align}
where $\alpha_m$ and $\beta_m$ are parameters, which are local to the dimension. 
\item \textsf{SelfCorrecting} (Self-correcting Point Process) \cite{isham1979self}: 
The intensity function of \textsf{SelfCorrecting} is assumed to increase steadily over time with the rate $\beta_m>0$; 
this trend is corrected by constant $\rho_m>0$ every time an event arrives.  
Its intensity function associated with dimension $m$ is given by 
\begin{align}
\lambda_m(t) = \exp{\big( \alpha_m + \beta_{m}(t-\rho_m N^m(t)) \big)}, 
\end{align}
where $\alpha_m$, $\beta_m$, and $\rho_m$ are parameters, and 
$N^m(t)$ is the number of events of dimension $m$ in $(0,t]$.  
\end{itemize}

\end{document}